# Defending Compute Thresholds Against Legal Loopholes

Matteo Pistillo        Pablo Villalobos

Table of Contents

**Executive Summary**

Existing legal frameworks on artificial intelligence ('AI') rely on training computational (or 'compute') thresholds as a proxy to identify potentially-dangerous AI models and trigger increased regulatory attention. In the United States, Section 4.2(a) of Executive Order 14110 instructs the Secretary of Commerce to require extensive reporting from developers of AI models above a certain training compute threshold. In the European Union, Article 51 of the AI Act establishes a presumption that AI models above a certain compute threshold have high impact capabilities and hence pose systemic risk, thus subjecting their developers to several obligations including capability evaluations, reporting, and incident monitoring. The bedrock of training compute thresholds lies in scaling laws—the observation that the compute used to train an AI model is correlated with its performance and, hence, its base capabilities.

In this paper, we examine some enhancement techniques that are capable of decreasing training compute usage while preserving, or even increasing, model capabilities. Since training compute thresholds rely on training compute as a metric and trigger for increased regulatory attention, these capability-enhancing and compute-saving techniques could constitute a legal loophole to existing training compute thresholds. In particular, we concentrate on four illustrative techniques—fine-tuning, model reuse (knowledge distillation, kickstarting, and reincarnation), model expansion, and above compute-optimal inference compute—with the goal of furthering the conversation about their implications on training compute thresholds as a legal mechanism and advancing policy recommendations that could address the relevant legal loopholes.

Our findings and recommendations can be summarized as follows.

- **Fine-tuning**
  - *Legal framework*: In Executive Order 14110, it is unclear whether the compute used for fine-tuning counts towards the training compute threshold.
  - *Potential loophole*: Developers could scale up AI models below the threshold through significant fine-tuning and circumvent the regulatory compliance set forth under Executive Order 14110.
  - *Policy recommendation*: Fine-tuning increases a large language model's pre-training performance by a statistically significant amount (~2%) once it reaches the threshold of 14.4% of the initial training compute. Hence, we recommend that, when fine-tuning exceeds 15% of the original training compute, the United States count fine-tuning towards meeting the training compute threshold set forth under Executive Order 14110.

- **Model reuse**
  - *Legal frameworks*: Executive Order 14110 and the AI Act do not clarify whether model reuse should be counted towards meeting the existing compute thresholds if the starting model (the 'teacher' model) falls below the threshold.
  - *Potential loopholes*: Developers might use these techniques to obtain AI models that: (i) have the same—or even better —capabilities than the starting model; (ii) do not cross the compute thresholds; (iii) would have crossed the compute thresholds without recurring to these techniques.
  - *Policy recommendation*: We recommend that the United States and the European Union consider a compute saving up to ~10x when calculating training compute thresholds for reused models.

- **Model expansion**
  - *Legal framework*: (i) It is unclear whether Executive Order 14110 calculates the compute used for model expansion towards meeting the compute threshold in cases in which the overall sum between the pre-training compute and the compute used for model expansion is greater than the compute threshold. Furthermore, (ii) Executive Order 14110 and the AI Act do not regulate cases in which both the starting and the resulting model are below the compute thresholds.
- *Potential loopholes*: In both cases, AI developers might scale up AI models through model expansion, circumventing regulatory compliance.
- *Policy recommendation*: We recommend that the United States and the European Union consider model expansion when calculating compute thresholds and factor a 20-76% compute saving through model expansion when this technique is used.

- **Above compute-optimal inference compute**
  - *Legal frameworks*: Executive Order 14110 and the AI Act do not consider inference compute for the purposes of calculating compute thresholds.
  - *Potential loophole*: AI developers might shift compute expenditures from training to inference and, within certain limits, obtain equally-capable models while avoiding regulatory compliance.
  - *Policy recommendation*: We recommend that the United States and the European Union consider inference compute when calculating compute thresholds, bearing in mind that: (a) an increase of 2-3 orders of magnitude ('OOM') above compute-optimal inference compute usually allows a saving of ~2 OOM in training compute; (b) for math and coding capabilities, an increase of 5-6 OOM above compute-optimal inference compute usually corresponds to a saving of 3-4 OOM in training compute.

# I. Introduction

The term 'compute thresholds' refers to a regulatory threshold devised for the governance of frontier AI models.[1] Compute thresholds offer a simple and clean, albeit imprecise,[2] solution to categorizing AI models and their potential for risk. Relying on scaling laws,[3] these thresholds aim at pre-selecting potentially-dangerous AI models by taking the compute used for their training as a proxy for their capabilities and potential ability to cause harm, and subjecting these models to greater regulatory attention including capability evaluations, reporting obligations, and incident monitoring.

At the time of writing, two compute thresholds have been successfully established: at the federal level in the United States through Executive Order 14110,[4] and in the European Union through the AI Act.[5] A third compute threshold came very close to existence at the state level in the United States, through California Senate Bill 1047.[6] The bill was approved by the California legislature but then vetoed by the State's Governor.[7] Despite reasonable criticism being leveled at compute thresholds,[8] it is possible that this type of regulatory threshold will become more prevalent across the globe in the coming years, given that it provides an 'easy' solution to categorize frontier AI models. For instance, the upcoming United Kingdom's frontier AI Bill might include a form of training compute threshold. Furthermore, the establishment of compute thresholds has been advocated for in the People's Republic of China by a group of influential scholars.[9]

---

[1] *See* Lennart Heim & Leonie Koessler, *Training Compute Thresholds: Features and Functions in AI Regulation*, ARXIV (last updated Aug. 6, 2024), https://doi.org/10.48550/arXiv.2405.10799; Matteo Pistillo et al., *The Role of Compute Thresholds for AI Governance*, GEORGE WASHINGTON JOURNAL OF LAW AND TECHNOLOGY (forthcoming, 2025).

[2] *See* Sara Hooker, *On the Limitations of Compute Thresholds as a Governance Strategy*, ARXIV (last updated July 30, 2024), https://doi.org/10.48550/arXiv.2407.05694; Pistillo et al., *supra* note 1, at 38 (describing the limits of compute thresholds).

[3] Scaling laws describe the predictable relationship between the inputs required to produce a model (compute, data, and parameters) and the resulting performance of the model. *See* Kaplan et al., *Scaling Laws for Neural Language Models*, ARXIV (Jan. 23, 2020), https://arxiv.org/abs/2001.08361.

[4] Executive Order No. 14110, Safe, Secure, and Trustworthy Development and Use of Artificial Intelligence, 88 F.R. 75191 (Oct. 30, 2023), https://www.whitehouse.gov/briefing-room/presidential-actions/2023/10/30/executive-order-on-the-safe-secure-and-trustworthy-development-and-use-of-artificial-intelligence/, at Sec. 4.2(b)–(c) [hereinafter 'Executive Order 14110'].

[5] Regulation (EU) 2024/1689 of the European Parliament and of the Council of 13 June 2024 laying down harmonized rules on artificial intelligence and amending Regulations (EC) No 300/2008, (EU) No 167/2013, (EU) No 168/2013, (EU) 2018/858, (EU) 2018/1139 and (EU) 2019/2144 and Directives 2014/90/EU, (EU) 2016/797 and (EU) 2020/1828, https://eur-lex.europa.eu/eli/reg/2024/1689/oj, at art. 51(2) [hereinafter 'AI Act']

[6] California Senate Bill No. 1047, Safe and Secure Innovation for Frontier Artificial Intelligence Systems Act (Feb. 7, 2024), at Sec. 22602(f) [hereinafter 'California Senate Bill 1047'].

[7] *See* Office of the Governor Gavin Newsom, *Governor Newsom announces new initiatives to advance safe and responsible AI, protect Californians* (Sep. 29, 2024), https://www.gov.ca.gov/2024/09/29/governor-newsom-announces-new-initiatives-to-advance-safe-and-responsible-ai-protect-californians/; Office of the Governor Gavin Newsom, *Veto message* (Sep. 29, 2024), https://www.gov.ca.gov/wp-content/uploads/2024/09/SB-1047-Veto-Message.pdf ("Key to the debate is whether the threshold for regulation should be based on the cost and number of computations needed to develop an AI model, or whether we should evaluate the system's actual risks regardless of these factors.").

[8] *See* Hooker, *supra* note 2.

**Executive Order 14110** requires the Secretary of Commerce—in consultation with the Secretary of State, the Secretary of Defense, the Secretary of Energy, and the Director of National Intelligence—to define within 90 days "the set of technical conditions" under which AI models must be subject to reporting obligations as set forth under Section 4.2(a).[10] Until then, the Secretary of Commerce must require that "any model that was trained using a quantity of computing power greater than 10^26 integer or floating-point operations" comply with Section 4.2(a) reporting requirements.[11] At the time of writing, no known AI model crosses the 1e26 OP or FLOP training compute threshold, which means that no existing model is subject to these proposed oversight measures through this pathway.[12]

The **AI Act** establishes that a general-purpose AI model poses systemic risk if "it has high-impact capabilities" as determined by "appropriate technical tools and methodologies, including indicators and benchmarks," or "a decision of the Commission."[13] In this respect, a general-purpose AI model is presumed to have high-impact capabilities if "the *cumulative* amount of computation used for its training measured in floating point operations is greater than 10^25."[14] In light of "evolving technological developments, such as algorithmic improvements or increased hardware efficiency," the Commission can amend the threshold through delegated acts, as well as "supplement benchmarks and indicators."[15] At the time of writing, around 20 models appear to cross the 10^25 FLOP threshold.[16]

The vetoed **California Senate Bill 1047** defined "covered model[s]" as AI models that: (i) were trained with a quantity of compute greater than 1e26 OP or FLOP, whose cost exceeded 100 million dollars;[17] or (ii) were created by fine-tuning a "covered model" with 3e25 FLOP or OP, or more, for a cost exceeding 10 million dollars.[18] The threshold was supposed to be updated by the Government Operations Agency on and after January 1, 2027.[19] The vetoed Bill also included the category of "[c]overed model derivative," which was defined as: (i) "[a]n unmodified copy of a covered model;" (ii) "[a] copy of a covered model" that has been "subjected to post-training modifications unrelated to fine-tuning;" (iii) "a copy of a covered model" that has been fine-tuned with *less than* 3e25 FLOP or OP (or the threshold to be determined by the Government Operations Agency on and after January 1, 2027), for a cost exceeding 10 million dollars; or (iv) "a copy of a covered model that has been combined with other software."[20]

---

[9] *See* Artificial Intelligence Law of the People's Republic of China (Draft for Suggestions from Scholars), CHINA LAW SOCIETY (Mar. 18, 2024), http://www.fxcxw.org.cn/dyna/content.php?id=26910, at art. 50(iii), art. 50–57, translated at Artificial Intelligence Law of the People's Republic of China (Draft for Suggestions from Scholars), CENTER FOR SECURITY AND EMERGING TECHNOLOGY (May 2, 2024), https://cset.georgetown.edu/publication/china-ai-law-draft/ ("Foundation models that have reached a certain level in aspects such as compute, parameters, or scale of use"); Matt Sheehan, X (Mar. 21, 2024), https://twitter.com/mattsheehan88/status/1770902104795729936.
[10] Executive Order 14110, *supra* note 4, at Sec. 4.2(b).
[11] Executive Order 14110, *supra* note 4, at Sec. 4.2(b).
[12] *See Large-Scale AI Models*, EPOCH (last updated Dec. 30, 2024; last consulted Dec. 30, 2024), https://epochai.org/data/large-scale-ai-models.
[13] AI Act, *supra* note 5, at Art. 51(1).
[14] *Id.*, at Art. 51(2).
[15] *Id.*, at Art. 51(3).
[16] *See Large-Scale AI Models*, *supra* note 12 (last consulted Dec. 30, 2024).

In this paper, we highlight four capability-enhancing techniques that might lead to legal loopholes in the mentioned training compute thresholds: (i) significant fine-tuning: (ii) model reuse, such as knowledge distillation, kickstarting, and reincarnation; (iii) model expansion; (iv) above-optimal inference compute use. These techniques enable AI developers to increase their models' capabilities while simultaneously conserving training compute resources, thereby potentially allowing them to circumvent the additional regulatory compliance triggered by the compute thresholds set forth under Section 4.2(a) of Executive Order 14110 and Article 55 of the AI Act. For instance, these loopholes can enable developers to elude the obligations to provide the U.S. Government with "information, reports, or records" on "the results of any developed dual-use foundation model's performance in relevant AI red-team testing,"[21] or to perform model evaluations, assess and mitigate systemic risks, collect information about serious incidents, and ensure adequate cybersecurity protection under the AI Act.[22] This fallacy in compute thresholds has been recently raised by other scholars, who observed that "many improvements dramatically improve performance but are currently completely ignored by compute thresholds since they don't contribute to training FLOP"[23] and "[t]he FLOPS threshold

---

[17] California Senate Bill 1047, *supra* note 6, at Sec. 22602.
[18] *Id*.
[19] *Id*.
[20] *Id*. (emphasis added). The final version of California Senate Bill 1047 replaced a previous draft, according to which covered models included AI models trained "using a quantity of computing power sufficiently large that it could reasonably be expected to have similar or greater performance as an artificial intelligence model trained using a quantity of computing power greater than 10^26 integer or floating-point operations in 2024 as assessed using benchmarks commonly used to quantify the general performance of state-of-the-art foundation models." *Cf.* the version of California Senate Bill 1047 dated May 16, 2024, https://leginfo.legislature.ca.gov/faces/billVersionsCompareClient.xhtml?bill_id=202320240SB1047&cversion=20230SB104794AMD.
[21] Executive Order 14110, *supra* note 4, at Sec. 4.2(a).
[22] AI Act, *supra* note 5, at Art. 55.
[23] *See* Hooker, *supra* note 2, at 13.

… incentivizes providers to optimize models to fall under the threshold without necessarily making those models less dangerous."[24]

## II. Legal Loopholes and Policy Suggestions

This Section examines four illustrative capability-enhancing and compute-saving techniques that might constitute legal loopholes to existing training compute thresholds.

- **II.A**: Fine-tuning above a certain threshold.
- **II.B**: Model reuse, such as knowledge distillation, kickstarting and reincarnation.
- **II.C**: Model expansion.
- **II.D**: Above-optimal inference compute use.

Each Section offers a description of:

**(i)** The techniques.
**(ii)** The potential legal loopholes to the existing training compute thresholds.
**(iii)** How the potential legal loopholes interact with the existing legal frameworks.[25]
**(iv)** Our policy recommendations to patch the described loopholes.

We believe that more research should be pursued on this topic. First, more techniques than the ones described in this paper are capable of enhancing the capabilities of trained models, including scaffolding, compression, and solution choice. Second, additional research could reach a more precise estimate of compute equivalent savings, as current estimates are based on observations from experiments that were not explicitly designed to measure them.

---

[24] Sandra Wachter, *Limitations and Loopholes in the EU AI Act and AI Liability Directives: What This Means for the European Union, the United States, and Beyond*, 26(3) YALE JOURNAL OF LAW & TECHNOLOGY (2024), at 698.
[25] We also consider how California Senate Bill 1047 would have interacted with the selected loopholes because, while it was vetoed and never entered into force, it could serve as a blueprint for future legislation in California, the United States, or in other countries.

## A. Fine-tuning

### (i) *Technique*

Large language models ('LLMs') are usually trained in at least two different stages—pre-training and fine-tuning. These stages rely on the same fundamental operations at the technical level, and are only distinguished by the amount of training, the type of data, and the effects. Pre-training is more computationally intensive and endows the model with powerful general-purpose capabilities. Fine-tuning is less demanding in terms of compute and relies on specially crafted data to refine a model's capabilities.[26]

Fine-tuning is a crucial component of modern LLMs, and most frontier models use some form of it.[27] Fine-tuning is generally used to adapt AI models to specific use-cases (like math[28]), to align AI models with human preferences,[29] and for instruction tuning—which is regarded as an essential process to elicit model capabilities.[30] As Figure 1 below shows, fine-tuning can have a significant impact on the capabilities of models, particularly in narrow tasks. However, it is generally accepted that fine-tuning in most cases does not significantly increase general capabilities, but merely modulates them.[31] Given its broad usefulness, large diffusion, and limited impact on general capabilities, fine-tuning is often considered as not warranting immediate monitoring or regulatory oversight.

---

[26] Kyle Miller & Andrew Lohn, *Techniques to Make Large Language Models Smaller: An Explainer*, CENTER FOR SECURITY AND EMERGING TECHNOLOGY (Oct. 11, 2023), https://cset.georgetown.edu/publication/techniques-to-make-large-language-models-smaller-an-explainer, at 3 ("Fine-tuning involves feeding more data to a trained model so it hones the ability to perform certain tasks … [T]his requires less compute and memory than the original training …"); Wenxin Jiang et al., *An Empirical Study of Pre-Trained Model Reuse in the Hugging Face Deep Learning Model Registry*, ARXIV (Mar. 5, 2023), https://doi.org/10.48550/arXiv.2303.02552, at 2 ("Through transfer learning, DNNs can be pre-trained on large datasets and fine-tuned to solve specialized tasks, leveraging a PTM's knowledge of one task to better teach it a similar task.").

[27] *See* OpenAI, *GPT-4 Technical Report*, ARXIV (last updated Mar. 4, 2024), https://doi.org/10.48550/arXiv.2303.08774 (RLHF); Google, *PaLM 2 Technical Report*, ARXIV (last updated Sep. 13, 2023), https://doi.org/10.48550/arXiv.2305.10403 (Instruction Tuning); Anthropic, *The Claude 3 Model Family: Opus, Sonnet, Haiku*, ANTHROPIC (Mar. 4, 2024), https://www-cdn.anthropic.com/de8ba9b01c9ab7cbabf5c33b80b7bbc618857627/Model_Card_Claude_3.pdf (Constitutional AI).

[28] *See* Ethan Dyer & Guy Gur-Ari, *Minerva: Solving Quantitative Reasoning Problems with Language Models*, GOOGLE RESEARCH (June 30, 2020), https://research.google/blog/minerva-solving-quantitative-reasoning-problems-with-language-models/ ("We show that by focusing on collecting training data that is relevant for quantitative reasoning problems, [] we achieve significant performance gains on a variety of difficult quantitative reasoning tasks.").

[29] In particular, reinforcement learning from human feedback (RLHF) or its variants, such as direct preference optimization (DPO), or reinforcement learning from AI feedback (RLAIF). *See* Long Ouyang et al., *Training language models to follow instructions with human feedback*, ARXIV (Mar. 4, 2022), https://doi.org/10.48550/arXiv.2203.02155, at 2 ("We focus on fine-tuning approaches to aligning language models. Specifically, we use reinforcement learning from human feedback to fine-tune GPT-3 to follow a broad class of written instructions.").

[30] *See* Shengyu Zhang et al., *Instruction Tuning for Large Language Models: A Survey,* ARXIV (Aug. 21, 2023), https://doi.org/10.48550/arXiv.2308.10792, at 1 ("instruction tuning [is] a crucial technique to enhance the capabilities and controllability of large language models.").

Figure 1 – Relationship Between Fine-tuning and Capabilities

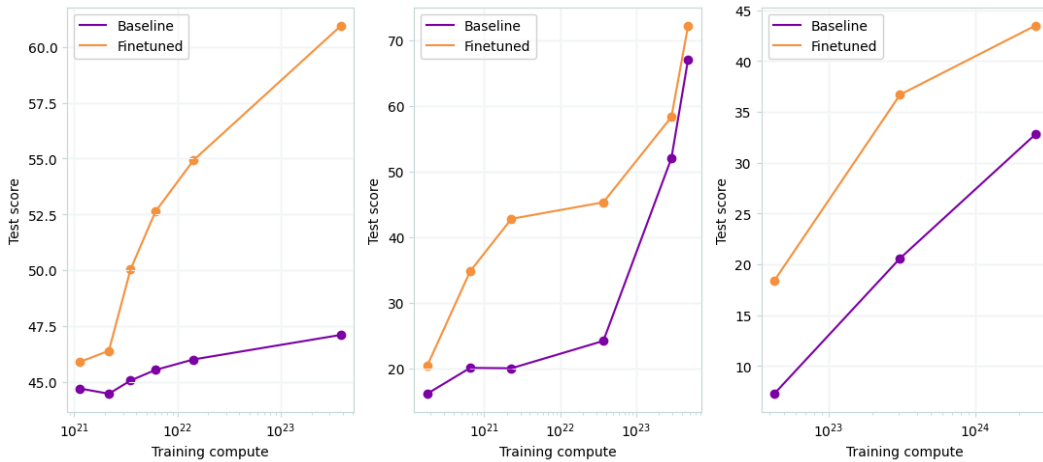

Training compute and downstream test scores for three families of fine-tuned models and their respective base models (left: Bloomz;[32] center: Flan;[33] right: Minerva[34]). Bloomz and Flan are instruction fine-tuned models, whereas Minerva is fine-tuned on a math dataset. In all cases, fine-tuning compute was smaller than 10% of training compute.

### *(ii)* <u>Potential Loophole</u>

Nonetheless, there is a risk that fine-tuning could circumvent existing compute thresholds. Consider the following example, visually represented in Figure 2 below. A developer has an existing AI model ('Model A'), which currently sits below the regulatory threshold. The developer wants to produce a new, more capable model using additional compute. If the developer trains a new model from scratch exceeding the threshold ('Model B'), it will be subject to regulatory compliance. The developer could then opt for a less-burdensome route: fine-tuning the original model (Model A) on general-purpose data with an additional amount of compute that is lower than the compute threshold, but that, if cumulated with the compute used for the initial training, would be greater than the compute threshold (or lower than the compute threshold but with comparable capabilities). This maneuver has the potential to exempt developers from regulatory compliance and possibly compromise the intended safety goals of the regulatory framework.

---

[31] *See* Nikhil Prakash et al., *Fine-Tuning Enhances Existing Mechanisms: A Case Study on Entity Tracking*, ARXIV (Feb. 22, 2024), https://doi.org/10.48550/arXiv.2402.14811, at 1 ("Our findings suggest that fine-tuning enhances, rather than fundamentally alters, the mechanistic operation of the model."); Samyak Jain et al., *Mechanistically analyzing the effects of fine-tuning on procedurally defined tasks*, ARXIV (Nov. 21, 2023), https://doi.org/10.48550/arXiv.2311.12786, at 1 ("fine-tuning rarely alters the underlying model capabilities [...] a minimal transformation [...] is typically learned on top of the underlying model capabilities.").

[32] *See* Niklas Muennighoff et al., *Cross Lingual Generalization through Multitask Finetuning*, ARXIV (Nov. 3, 2021), https://doi.org/10.48550/arXiv.2004.14203.

[33] *See* Jason Wei et al., *Finetuned Language Models Are Zero-Shot Learners*, ARXIV (Sep. 3, 2021), https://doi.org/10.48550/arXiv.2109.01652.

[34] *See* Aitor Lewkowycz et al., *Solving Quantitative Reasoning Problems with Language Models*, ARXIV (Jun. 29, 2022), https://doi.org/10.48550/arXiv.2206.14858.

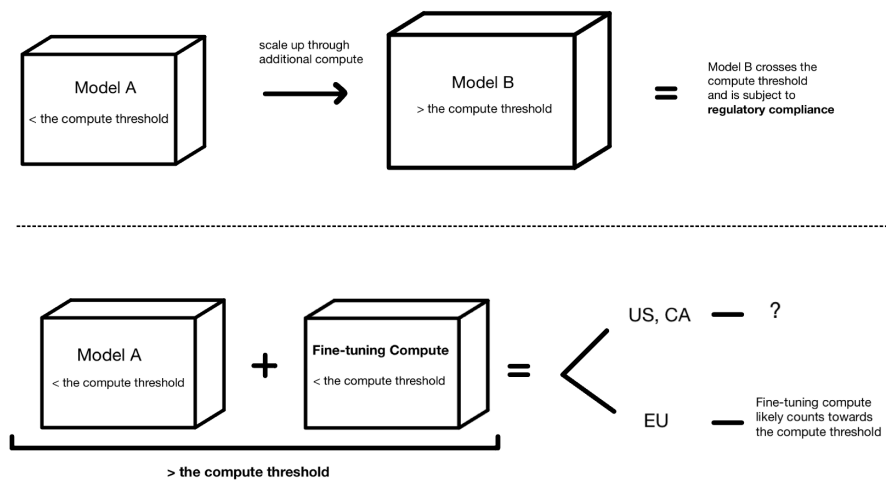

Figure 2 – Fine-tuning as a Potential Loophole

*(iii)* *Legal Frameworks*

**(a)** *Executive Order 14110*
Executive Order 14110 does not contain any specification around how fine-tuning should be considered for the purposes of the compute threshold set forth under Section 4.2(a)—making it unclear whether and to what extent fine-tuning compute counts towards the 1e26 OP or FLOP compute thresholds set forth therein (*see* Figure 2 above).

**(b)** *AI Act*
Article 51(2) of the AI Act specifies that the 1e25 FLOP compute threshold—which triggers a presumption of systemic risk—includes the "*cumulative* amount of computation used for [] training."[35] Recital 111 of the AI Act clarifies that "[t]he cumulative amount of computation used for training" includes "the computation used *across the activities and methods that are intended to enhance the capabilities of the model prior to deployment*, such as pre-training, synthetic data generation and *fine-tuning*."[36] Therefore, in the European Union, fine-tuning compute likely counts towards the 1e25 FLOP compute threshold (*see* Figure 2 above).[37]

---

[35] AI Act, *supra* note 5, at Art. 51(2) (emphasis added) ("A general-purpose AI model shall be presumed to have high impact capabilities pursuant to paragraph 1, point (a), when the cumulative amount of computation used for its training measured in floating point operations is greater than 10^25.").

**(c)** *California Senate Bill 1047*

Under the vetoed California Senate Bill 1047, the definition of 'covered model' would have included AI models that were "created by fine-tuning a covered model using a quantity of computing power equal to or greater than three times 10^25 integer or floating-point operations."[38] In other words, covered models would have included models above 1e26 OP or FLOP[39] that were then fine-tuned with 3e25 OP or FLOP or more. If a developer had retrained a covered model (above 1e26 OP or FLOP) with less than 3e25 OP or FLOP, the model would have fallen within the category of 'covered model derivatives.'[40] California Senate Bill 1047 defined fine-tuning as "adjusting the model weights of a trained covered model or covered model derivative by exposing it to additional data."[41] Therefore, fine-tuning would have counted towards compute thresholds in California, but *only if* the starting model—i.e., the model before fine-tuning—had already exceeded 1e26 OP or FLOP. Whereas it appears that fine-tuning would not have counted towards compute thresholds if the starting model had not already been a 'covered model,' such as for instance if the starting model was at 1e25 OP or FLOP (*see* Figure 2 above).

*(iv)  Policy Recommendation*

The United States could still patch this potential legal loophole. As mentioned above, the Secretary of Commerce has yet to define the "set of technical conditions" for models that would be subject to the reporting requirements.[42] Compute thresholds might also be subject to judicial review at a later point, which could offer courts an opportunity to interpret the gaps.

---

[36] *Id.*, at Recital 111 (emphasis added).
[37] Nonetheless, it remains unclear whether the term 'cumulative' includes any and all fine-tuning compute, or only fine-tuning compute above a certain threshold of relevance.
[38] California Senate Bill 1047, *supra* note 6, at Sec. 22602.
[39] *Id.* ("An artificial intelligence model trained using a quantity of computing power greater than 10^26 integer or floating-point operations, the cost of which exceeds one hundred million dollars ($100,000,000) when calculated using the average market prices of cloud compute at the start of training as reasonably assessed by the developer.").
[40] *Id.* ("a quantity of computing power not exceeding three times 10^25 integer or floating point operations.").
[41] *Id.*
[42] Executive Order 14110, *supra* note 4, at Sec. 4.2(b).

Adding a specification that the 1e26 OP or FLOP threshold is 'cumulative,' along the lines of the AI Act, is certainly one option for making the compute threshold set forth under Executive Order 14110 more comprehensive. However, there may be a better alternative. In fact, the expression 'cumulative' might include within the threshold any and all compute used to fine-tune AI models, including for the purposes of alignment with human preferences and capability elicitation. In addition to being undesirable, the compute necessary to fine-tune an AI model and the relative costs are only a very small fraction of training compute—typically < 1% and sometimes < 0.01% of the pre-training cost.[43] This makes fine-tuning very widespread and thus impractical for regulators to track. Consider, for instance, that Hugging Face's model hub lists more than one million models, of which more than 80,000 are fine-tuned or modified versions of Llama.[44]

A policy solution targeting fine-tuning should, therefore, distinguish legitimate uses of fine-tuning from disguised model retraining. Since fine-tuning rarely alters an AI model's general capabilities,[45] one could distinguish the two based on the generation of new general capabilities.[46] In this way, the demarcation line would be set at the minimum amount of compute necessary to create new capabilities—in other words, at the minimum detectable change. When fine-tuning is so extensive that an AI model develops new capabilities, it should be counted within the compute threshold. Based on the data from the *Chinchilla* paper, a 14.4% increase in training compute is sufficient to improve an AI model's test loss by 2%, a statistically significant improvement.[47] Therefore, the minimum detectable change sits around 14.4% of training compute—which can be rounded up to 15%, to avoid false positives. With a 15% increase in training compute, either in one solution or in multiple installments, one can confidently increase capabilities by at least 1%.

---

[43] Tom Davidson et al., *AI capabilities can be significantly improved without expensive retraining*, ARXIV (Dec. 12, 2023), https://doi.org/10.48550/arXiv.2312.07413, at 2.

[44] *See Models*, HUGGING FACE, https://huggingface.co/models (last consulted Dec. 30, 2024).

[45] *See* Samyak Jain et al., *Mechanistically analyzing the effects of fine-tuning on procedurally defined tasks*, ARXIV (Aug. 21, 2024), https://doi.org/10.48550/arXiv.2311.12786, at 1 ("fine-tuning rarely alters the underlying model capabilities."); Miller & Lohn, *supra* note 26, at 4 ("the performance of these models is generally poorer than large, proprietary models.").

[46] It is difficult to define what amounts to a 'new capability.' Scholars have adopted different approaches. There might also be exceptions to the general rule that small amounts of fine-tuning compute are not consequential for the creation of new capabilities. *Cf.* Lee Sharkey et al., *A Causal Framework for AI Regulation and Auditing*, APOLLO RESEARCH (Nov. 8, 2023), https://www.apolloresearch.ai/research/a-causal-framework-for-ai-regulation-and-auditing ("Although some fine-tuning updates may have only small changes to the absolute capabilities of a system, some fine-tuning approaches may contribute significant absolute capabilities with only small numbers of updates. For example, using the UL2R mixture of training objectives to finetune a language model, it is possible to drastically improve performance on downstream tasks using only 0.1% to 1% of the pretraining computation costs.").

[47] *See* Jordan Hoffmann et al., *Training Compute-Optimal Large Language Models*, ARXIV (Mar. 29, 2022), https://doi.org/10.48550/arXiv.2203.15556. Based on the data available in the paper, the standard deviation of the loss is around 1%. Therefore, in order to achieve a 95% certainty that fine-tuning is improving capabilities, we need a decrease in test loss of around 2%. This corresponds to $98\%^{(-1/0.15)} = 14.4\%$ increase in training compute, using the *Chinchilla* scaling law.

In light of the above, we suggest that **fine-tuning compute is counted towards the existing compute threshold when—in one instance or in the aggregate—it is greater than 15% of the compute used to train the AI model**. By contrast, fine-tuning compute should not be counted whenever it is lower. The effect of this policy recommendation would be to also include within the existing thresholds AI models that are developed with a lower amount of initial training compute and that are subsequently fine-tuned with compute exceeding 15% of the training compute. For instance, assuming a fine-tuning threshold at 15% of training compute, AI models developed with 1.5e25 OP or FLOP would fall within the threshold established in the United States.

This policy suggestion would be desirable for three reasons. First, it emphasizes fairness. AI models with the same capabilities would be treated equally, regardless of the capacity-enhancing technique used. At the same time, this solution utilizes and relies on the compute threshold already identified in Executive Order 14110, at 1e26 OP or FLOP. It does not aim to expand this threshold to include AI models with lower capabilities than the frontier AI models already identified as requiring additional oversight. Instead, it only seeks to explain how this threshold could be better applied in a context in which AI developers utilize extensive fine-tuning. Second, enactment and enforcement of this policy solution would be practical for regulators. At the time of writing, less than 10 developers appear to have developed AI models above 1.5e25 OP or FLOP globally,[48] making enforcement of this policy—including the relevant monitoring—feasible for regulators. Third, regulatory compliance would not burden individuals, small organizations, or research centers that fine-tune models. The cost to train a model at 1e26 FLOP or OP is estimated around 100 million USD[49]—which means that the cost of fine-tuning an AI model with compute above 15% of the model's initial training compute (over 15 million USD) would likely be sustainable only for large organizations.

Our policy suggestion would be best complemented by reporting requirements triggered once fine-tuning compute crosses the 15% threshold. For instance, Article 52 of the AI Act—according to which "the relevant provider shall notify the Commission without delay and in any event within two weeks after that requirement" (i.e., the crossing of the 1e25 FLOP training compute threshold) "is met or it becomes known that it will be met"—could be adapted to include fine-tuning compute above the relevant threshold.[50]

---

[48] *See Large-Scale AI Models*, *supra* note 12 (last consulted Dec. 30, 2024). We note that for many models FLOP estimates are not publicly available. Hence, the overall number could be higher.

[49] *See* Ben Cottier et al., *How Much Does It Cost to Train Frontier AI Models?*, EPOCH AI (Jun. 3, 2024), https://epochai.org/blog/how-much-does-it-cost-to-train-frontier-ai-models; *Key Trends and Figures in Machine Learning*, EPOCH AI, https://epochai.org/trends#investment (estimating the total cost of developing Google DeepMind's Gemini Ultra around 130 million dollars).

## B.     Model Reuse: Knowledge Distillation, Kickstarting, and Reincarnation

### *(i)   Techniques*

Knowledge distillation, kickstarting and reincarnation are three techniques that fall within the realm of model reuse.[51] The term 'model reuse' refers to the practice of "reus[ing] pre-trained models to help further model building."[52] In other words, "rather than building a model from scratch," AI developers "construct a model by utilizing existing available models, mostly trained for other tasks."[53]

**Knowledge distillation** consists in having a newer, smaller AI model learn from an older, larger AI model or ensemble of AI models.[54] A larger model or multiple small models generate data to train a new model, which will then have the same capabilities as those initial models. In other words, knowledge distillation distills knowledge from "a large teacher model" into a "small student model."[55] The student model "learns from the outputs of an old teacher model."[56] Rather than developing two "cumbersome model[s]" from scratch, the second AI model can learn from the first. In this way, AI developers can improve "smaller and cheaper models by distilling knowledge from expensive and proprietary models, such as those behind ChatGPT."[57]

---

[50] AI Act, *supra* note 5, at Art. 52 and Recital 112 ("The provider should notify the AI Office at the latest two weeks after the requirements are met or it becomes known that a general-purpose AI model will meet the requirements that lead to the presumption. This is especially relevant in relation to the threshold of floating point operations because training of general-purpose AI models takes considerable planning which includes the upfront allocation of compute resources and, therefore, providers of general-purpose AI models are able to know if their model would meet the threshold before the training is completed.").

[51] This realm includes more model reuse techniques. We focus on these three for illustrative purposes.

[52] Peng Zhao et al., *Handling Concept Drift via Model Reuse*, ARXIV (Sep. 8, 2018), https://doi.org/10.48550/arXiv.1809.02804, at 3 ("Model Reuse is an important learning problem, also named as model transfer, hypothesis transfer learning, or learning from auxiliary classifiers. The basic setting is that one desires to reuse pre-trained models to help further model building."). *See also* Yujie Ji et al., *Model-Reuse Attacks on Deep Learning Systems*, ARXIV (Dec. 2, 2018), https://doi.org/10.48550/arXiv.1812.00483, at 1 ("Today's machine learning (ML) systems are large, complex software artifacts. Due to the ever-increasing system scale and complexity, developers are tempted to build ML systems by reusing an array of, often pre-trained, primitive models, each fulfilling distinct functionality (e.g., feature extraction)."); Wenxin Jiang et al., *An Empirical Study of Pre-Trained Model Reuse in the Hugging Face Deep Learning Model Registry*, ARXIV (Mar. 5, 2023), https://doi.org/10.48550/arXiv.2303.02552, at 2, 4 ("Most interview participants take PTMs from model registries and apply transfer learning techniques to the model. They either "fine-tune an existing PTM" by (optionally) extending architecture and training on a task-specific dataset, or 'build a new model on top of the pre-trained one'.").

[53] Yang Yang et al., *Deep Learning for Fixed Model Reuse*, 31(1) PROCEEDINGS OF THE AAAI CONFERENCE ON ARTIFICIAL INTELLIGENCE (Feb. 13, 2017), https://doi.org/10.1609/aaai.v31i1.10855, at 2831-2832, 2836.

[54] Geoffrey Hinton et al., *Distilling the Knowledge in a Neural Network*, ARXIV (Mar. 9, 2015), https://doi.org/10.48550/arXiv.1503.02531, at 1 (explaining how, once "an ensemble of separately trained models or a single very large model" has been trained, knowledge can be transferred from such model to a small 'student' model).

**Kickstarting** "employ[s] already trained agents as teachers to share demonstrations with a student who has to learn the same task from scratch."[58] However, in contrast to distillation, the kickstarted 'student' AI model is allowed to surpass its 'teacher' in performance.[59]

**Reincarnation** refers to techniques in reinforcement learning that leverage previously trained agents to quickly train a new model, instead of training it from scratch.[60]

### *(ii)* *Potential Loophole*

All these techniques of model reuse—knowledge distillation, kickstarting, and reincarnation—constitute important scientific development. However, they could also represent a legal loophole to existing training compute thresholds as they can enable a significant reduction of compute consumption by exploiting already-existing models[61] and expedited learning in real-world tasks.[62] These techniques "leverage existing computational work"[63] so that the "[t]raining compute expended on earlier models can partially substitute for the compute needed in subsequent training runs,"[64] thus cutting down computational costs.[65] In other words, model reuse aims at increasing 'adjusted compute,' which has been defined as the "hypothetical amount of compute that would be required to train a model from scratch to reach the same level of performance that the actual model reached during training while employing model recycling."[66]

---

[55] Jianping Gou et al., *Knowledge Distillation: A Survey*, ARXIV (last updated May 20, 2023), https://doi.org/10.48550/arXiv.2006.05525, at 1-2. *See also* Miller & Lohn, *supra* note 26, at 3 ("One technique is to "distill" a subset of a large model's superior knowledge and capabilities into a smaller model, analogous to how an older teacher transfers some of their knowledge to a younger student.")

[56] Matthew Barnett, *The Limited Benefit of Recycling Foundation Models*, EPOCH AI (July 7, 2023), https://epochai.org/blog/the-limited-benefit-of-recycling-foundation-models.

[57] Miller & Lohn, *supra* note 26, at 3.

[58] Parham Gohari et al., *Privacy-Preserving Kickstarting Deep Reinforcement Learning with Privacy-Aware Learners*, ARXIV (last updated June 4, 2021), https://doi.org/10.48550/arXiv.2102.09599, at 1.

[59] Simon Schmitt et al., *Kickstarting deep reinforcement learning*, ARXIV (Mar. 10, 2018), https://doi.org/10.48550/arXiv.1803.03835.

[60] Rishabh Agarwal et al., *Reincarnating Reinforcement Learning: Reusing Prior Computation to Accelerate Progress*, ARXIV (last updated Oct. 4, 2022), https://doi.org/10.48550/arXiv.2206.01626.

[61] Yang Yang et al., *Deep Learning for Fixed Model Reuse*, 31(1) PROCEEDINGS OF THE AAAI CONFERENCE ON ARTIFICIAL INTELLIGENCE (Feb. 13, 2017), https://doi.org/10.1609/aaai.v31i1.10855, at 2831-2832, 2836.

[62] Yi-Kai Zhang et al., *ZhiJian: A Unifying and Rapidly Deployable Toolbox for Pre-trained Model Reuse*, ARXIV (Aug. 17, 2023), https://doi.org/10.48550/arXiv.2308.0915, at 1.

[63] Agarwal et al., *supra* note 60, at 2.

[64] Barnett, *supra* note 56.

[65] Wenxin Jiang et al., *An Empirical Study of Pre-Trained Model Reuse in the Hugging Face Deep Learning Model Registry*, ARXIV (Mar. 5, 2023), https://doi.org/10.48550/arXiv.2303.02552, at 2.

Figure 3 – Relationship Between Model Reuse and Capabilities

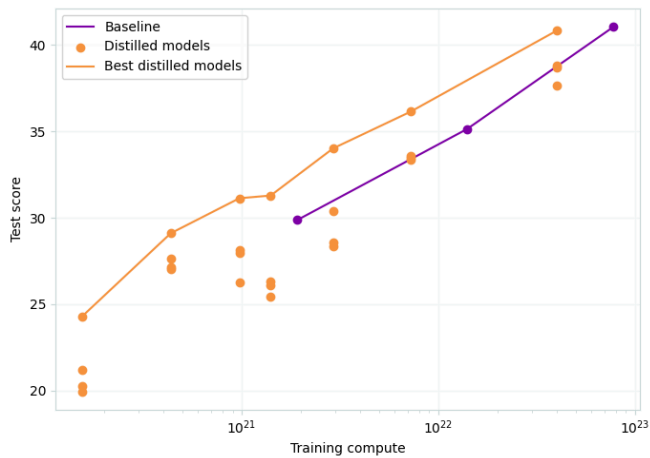

Performance of baseline and distilled models as a function of training compute. The best distilled models can achieve the same performance as non-distilled models using 60% less compute.[67]

As shown in Figure 3 above, compute savings from model reuse can be substantial.

A **distilled model** can be *up to* ~10x smaller than its teacher. For instance, Orca learned from GPT-4 and ChatGPT. Its size is only 13B parameters,[68] and its compute is 10x smaller than that of ChatGPT.[69] BERT's knowledge has been distilled into Tiny-BERT, a smaller "student" model that is 7.5x smaller than BERT.[70] DistilBERT reduced BERT's size by 40%.[71] MiniMA was distilled from LLaMA2-7B using 50x less compute.[72]

---

[66] Barnett, *supra* note 56 (noting that "[t]he value of model recycling is measured by the ratio of adjusted compute to real compute A/C, which is beneficial when A/C > 1").

[67] Data from Yuxian Gu et al., *MiniLLM: Knowledge Distillation of Large Language Models*, ARXIV (June 14, 2023), https://doi.org/10.48550/arXiv.2306.08543.

[68] *See* Subhabrata Mukherjee et al., *Orca: Progressive Learning from Complex Explanation Traces of GPT-4*, ARXIV (Jun. 5, 2023), https://doi.org/10.48550/arXiv.2306.02707.

[69] *See* Tom Davidson et al., *AI capabilities can be significantly improved without expensive retraining*, ARXIV (Dec. 12, 2023), https://doi.org/10.48550/arXiv.2312.07413, at 17.

[70] Xiaoqi Jiao et al., *TinyBERT: Distilling BERT for Natural Language Understanding*, ARXIV (last updated Oct. 16, 2020), https://doi.org/10.48550/arXiv.1909.10351, at 1.

[71] Victor Sanh et al., *DistilBERT, a distilled version of BERT: smaller, faster, cheaper and lighter*, ARXIV (Mar. 1, 2020), https://doi.org/10.48550/arXiv.1910.01108, at 5.

[72] *See* Chen Zhang et al., *Towards the Law of Capacity Gap in Distilling Language Models*, ARXIV (Jul. 25, 2024), https://doi.org/10.48550/arXiv.2311.07052, at 3.

Similarly, it has been estimated that a **kickstarted AI model** can match the performance of an AI model trained from scratch in ~10x fewer steps.[73]

With respect to **reincarnated models**, scholars have shown that the compute savings are 10x-15x for matching the performance of the previous model, and 2x-5x for surpassing it.[74]

If we assume that distilled, kickstarted, and reincarnated models can be ~10x smaller than their teachers, it means that AI developers can use up to roughly ~10x less compute to develop an AI model with capabilities comparable to the teacher model, or even better capabilities.

Figure 4 – Model Reuse as a Potential Loophole

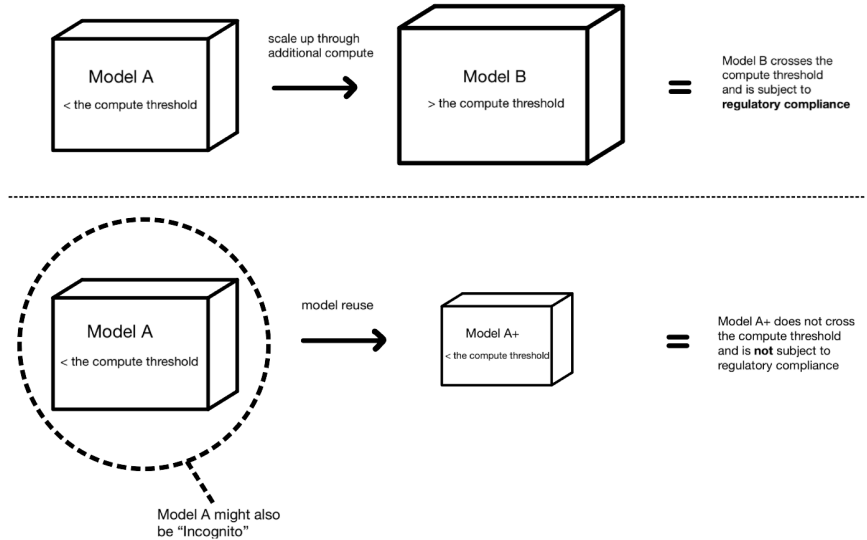

Therefore, AI developers interested in scaling up an existing model that does not yet surpass an existing compute thresholds ('Model A'), could either: (i) develop a new model ('Model B') using more compute, thereby crossing the compute threshold and ensuring the model be subject to regulatory compliance; or (ii) use knowledge distillation, kickstarting, or reincarnation on an existing model *below* the threshold (Model A) and potentially obtain an AI model ('Model A+') that has the same capabilities of the scaled-up model (Model B) but with a computational expenditure that falls below the compute threshold. As shown in Figure 4 above, this second

---

[73] Simon Schmitt et al., *Kickstarting Deep Reinforcement Learning*, ARXIV (Mar. 10, 2018), https://doi.org/10.48550/arXiv.1803.03835, at 1 ("We show that by kickstarting with multiple, task-specific expert teacher agents, we realize significant gains, with the kickstarted agent matching the performance of an agent trained from scratch in 9.58× fewer steps, and surpassing its final performance by 42.2%.").

[74] Agarwal et al., *supra* note 60, at 2.

route would enable AI developers to avoid regulatory compliance triggered by compute thresholds.

A third option for AI developers could be to: (iii) develop a 'teacher model' *above* the compute threshold, and never launch it on the market ('Incognito' Model A); then use knowledge distillation, kickstarting, or reincarnation to obtain a marketable AI model that sits below the threshold (Model A+). Similarly to option (ii), this third option could spare AI developers the burden of threshold-triggered regulatory compliance, as shown in Figure 4 above. However, this route heavily depends on how thoroughly compute reporting is verified by regulatory authorities and whether they check for AI models that are not deployed externally.

*(iii) Legal Frameworks*

**(a)** *Executive Order 14110*
The Executive Order on AI does not address model reuse.

**(b)** *AI Act*
Similarly, model reuse is not addressed by the AI Act.

**(c)** *California Senate Bill 1047*
The proposed definition of 'covered model derivative' included "a copy of a covered model that had been subjected to *post-training modifications unrelated to fine-tuning*,"[75] or "that had been combined with other software."[76] This could have included model reuse. However, California Senate Bill 1047 only considered post-training modifications applied to 'covered models.' Therefore, reused AI models below the threshold (1e26 OP or FLOP) would not have been regulated.

*(iv) Policy Recommendations*

Two policy solutions could patch this legal loophole. First, the model derived through knowledge distillation, kickstarting, or reincarnation (Model A+) could be **subject to regulatory compliance every time the starting 'teacher' model** (Model A)—whether incognito or not—**is greater than the compute threshold**, regardless of whether Model A+ sits below or above the compute threshold.

Second, when knowledge distillation, kickstarting, or reincarnation are used, the **relevant compute thresholds could be lower by a number of times equal to the compute savings**.

---

[75] California Senate Bill 1047, *supra* note 6, at Sec. 22602 (emphasis added).
[76] *Id.*

Assuming that these techniques allow AI developers to obtain AI models with up to ~10x less compute but equal or even better capabilities, existing compute thresholds would therefore be adjusted to 1e25 OP or FLOP in the United States, and 1e24 FLOP in the European Union. In other words, the compute threshold would be ~10x smaller when these techniques are employed. These policy solutions would ensure that reused AI models (Model A+) are still subject to regulatory compliance triggered by compute thresholds given that, had these techniques not been used, they would have crossed the compute threshold (Model B).

At the time of writing, enforcing this policy approach would not be burdensome for country regulators given that less than 20 organizations have developed AI models above 1e25 OP or FLOP and less than 40 above 1e24 FLOP or OP.[77] No existing AI model crosses the 1e26 OP or FLOP threshold.[78] In other words, there are relatively few AI developers that would be subject to regulatory oversight even if these policy solutions were adopted.[79]

## C. Model Expansion

### *(i) Technique*

Model expansion refers to a technique in which "a smaller model's pre-trained parameters are used to initialize a subset of a larger model's parameters,"[80] either by "deriving new neurons from the existing ones" or "initializing new parameters separately."[81] In other words, new learnable parameters are added to an already-trained AI model to make it larger and more capable. Simplifying the concept with a metaphor, model expansion is comparable to enhancing a human brain by adding additional gray matter that contains no knowledge and is ready to be trained. The larger model learns faster than a model of the same size trained from scratch by virtue of having parts of it already trained.

---

[77] *See Large-Scale AI Models, supra* note 12 (last consulted Dec. 30, 2024).

[78] *Id.* (last consulted Dec. 30, 2024).

[79] On the other hand, however, it should be noted that these techniques are relatively cost effective. It is therefore possible that the number of developers using these techniques could increase significantly. For instance, Orca was developed by a small team of six. *See* Subhabrata Mukherjee et al., *Orca: Progressive Learning from Complex Explanation Traces of GPT-4*, ARXIV (June 5, 2023), https://doi.org/10.48550/arXiv.2306.02707.

[80] Peihao Wang et al., *Learning to Grow Pretrained Models for Efficient Transformer Training*, ARXIV (Mar 2, 2023), https://doi.org/10.48550/arXiv.2303.00980, at 1.

[81] Wenyu Du et al., *Stacking Your Transformers: A Closer Look at Model Growth for Efficient LLM Pre-Training*, ARXIV (May 24, 2024), https://doi.org/10.48550/arXiv.2405.15319, at 1, 4.

*(ii)    Potential Loophole*

As shown in Figure 5 below, model expansion enables considerable compute savings. Such savings are estimated in a range between 20 and 76%, depending on the exact technique used.[82]

Figure 5 – Relationship Between Model Expansion and Capabilities

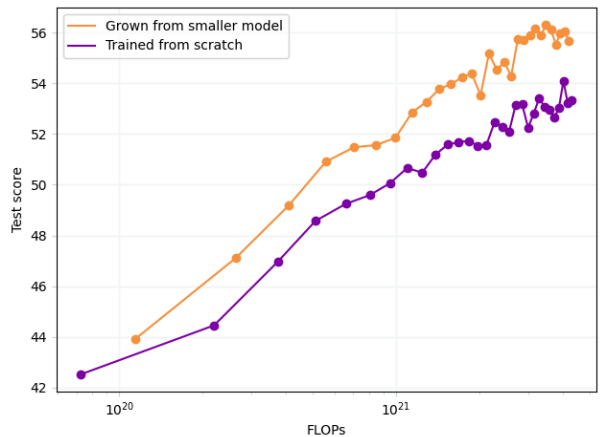

Comparison of the performance of models trained from scratch and grown from a smaller previously trained model. The grown models reach the same performance in ~50% less compute than what a regular training run would require.[83] Training the small model required 1% of the total compute.

Hence, instead of taking an AI model below the compute threshold ('Model A') and scaling it up above the threshold ('Model B'), AI developers could add new untrained modules to Model A and train only these new parts with a quantity of compute that is lower than the compute thresholds. This process could lead to two potential loopholes, visually represented in Figure 6 below.

---

[82] Peihao Wang et al., *Learning to Grow Pretrained Models for Efficient Transformer Training*, ARXIV (Mar. 2, 2023), https://doi.org/10.48550/arXiv.2303.00980, at 2 ("LiGO" "(learned linear growth operator (LiGO)" "saves 44.7% and 22.5% FLOPs for training BERT-Base and GPT2-Medium from scratch by reusing pretrained smaller models that are half as big."); Yiqun Yao et al., *Masked Structural Growth for 2x Faster Language Model Pre-training*, ARXIV (May 4, 2023), https://doi.org/10.48550/arXiv.2305.02869, (showing that Masked Structural Growth (MSG) is "significantly faster than related work," being able to "achieve up to 2.2x speedup in pre-training different types of language models while maintaining comparable or better downstream performances."); Sheng Shen et al., *Staged Training for Transformer Language Models*, ARXIV (Mar. 11, 2022), https://doi.org/10.48550/arXiv.2203.06211, at 7-8, 10 ("Empirical evaluations show up to 22% compute saving."); Longwei Zou et al., *A Multi-Level Framework for Accelerating Training Transformer Models*, ARXIV (Apr. 7, 2024), https://doi.org/10.48550/arXiv.2404.07999, at 8-9 (showing compute savings up to ~50% (51.6%)); Yu Pan, *Reusing Pretrained Models by Multi-linear Operators for Efficient Training*, ARXIV (Oct. 16, 2023), https://doi.org/10.48550/arXiv.2310.10699, at 9 ("Compared to baselines, Mango can achieve the highest savings in FLOPs with 76.4% for DeiT-B, 39.2% for BERT-Base, and 59.9% for GPT-Base from the Scratch model."); Abhishek Panigrahi et al., *Efficient Stagewise Pretraining via Progressive Subnetworks*, ARXIV (Feb. 8, 2024), https://doi.org/10.48550/arXiv.2402.05913, at 2 ("On BERT-Base, RAPTR demonstrates notable improvements in the pretraining loss when compared to gradual stacking at similar FLOPs, while being better than baseline training with 33% fewer FLOPs.").

[83] *Id.*, at 25.

I. The resulting model could have capabilities that are equivalent to Model B, while using an amount of compute that is *smaller* than the compute thresholds. This could allow AI developers to avoid the regulatory compliance triggered by training compute thresholds.

II. While the training compute and the growth compute are each smaller than the compute thresholds, the resulting model might *exceed* compute thresholds in the aggregate. Depending on how countries consider the compute used for model expansion, this could also allow AI developers to elude regulatory compliance.

Figure 6 – Model Expansion as a Potential Loophole

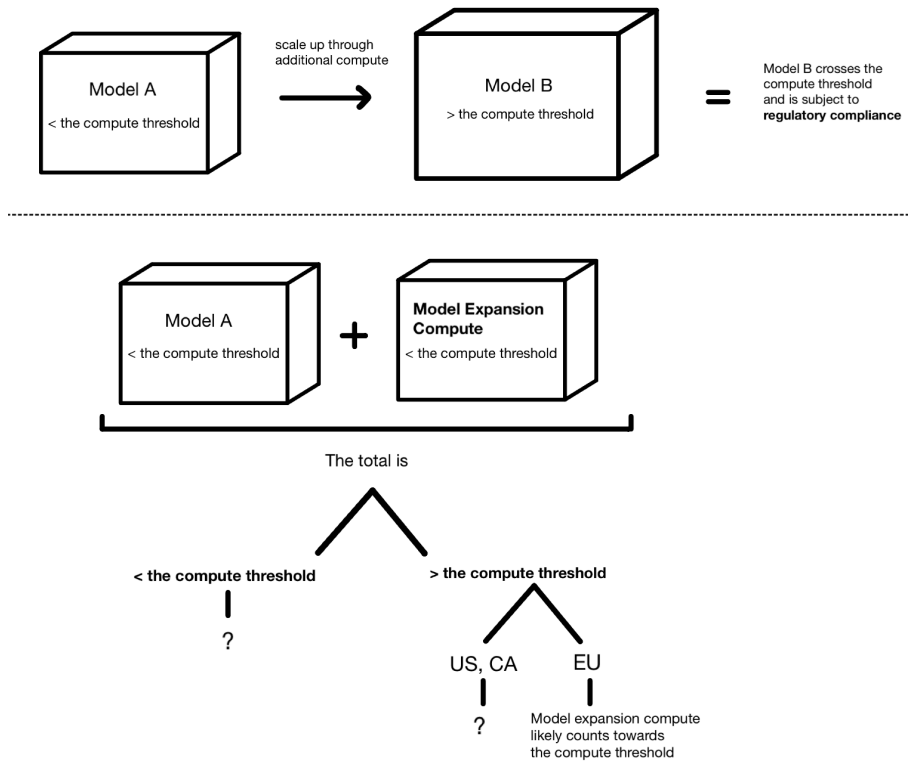

*(iii)* <u>*Legal frameworks*</u>

**(a)** *Executive Order 14110*
Executive Order 14110 does not regulate the first potential loophole described above (I). Similarly, with respect to the second potential loophole described above (II), it is unclear whether the compute used for model expansion counts towards the compute thresholds established under Executive Order 14110 (*see* Figure 6 above).

**(b)** *AI Act*
The AI Act does not regulate the first potential loophole described above (I). Instead, with respect to the second potential loophole described above (II), both the training compute and the compute used for model expansion would likely be counted towards the "cumulative" compute threshold in the European Union and the resulting model would thus likely fall within the scope of compute thresholds in the European Union (*see* Figure 6 above).

**(c)** *California Senate Bill 1047*
The California Senate Bill 1047 does not regulate the first potential loophole described above (I). With respect to the second potential loophole (II), California Senate Bill 1047 would have only regulated cases in which: (i) "post-training modifications unrelated to fine-tuning"—such as model expansion—are applied to "[a] copy of a covered model;" or (ii) "other software" is combined with "[a] copy of a covered model." If the starting model (Model A) were not already a 'covered model'—i.e., it fell below the compute threshold—then model expansion would not have been regulated by the proposed California Senate Bill 1047 compute threshold (*see* Figure 6 above).

*(iii)* <u>*Policy Recommendations*</u>

Following the rationale already laid out with respect to fine-tuning and model reuse (*see* previous Sections II.A-B), a possible solution to the model expansion loophole could be to factor the relevant compute savings in the calculation of compute thresholds. Assuming a compute saving in the range of 20-76%, AI developers could be subject to the regulatory compliance triggered by compute thresholds when: (a) they use model expansion; and (ii) the resulting AI model crosses 5e25 OP or FLOP in the United States (or 5e24 FLOP in the European Union). For a conservative approach that takes into account a possible 76% compute saving, a model expansion threshold could even be set at 2e25 OP or FLOP in the United States (or 2e24 FLOP in the European Union).

## D. Inference

### *(i)* *Technique*

Inference is the process through which a trained AI model is used to make new predictions based on users' input,[84] such as in the form of a prompt.[85] Each of the users' requests entails a compute expenditure, which can be defined as 'inference compute per request.'[86] When we refer to inference compute in this paper, we mean inference compute per request.

The current common practice is to train AI models with a dataset size proportional to the number of parameters.[87] The compute per inference token is proportional to the model size. Hence, in a model trained compute-optimally,[88] inference compute is proportional to—or slightly less than—the square root of the training compute.[89] For example, the compute-optimal inference compute for a model trained on 1e24 OP or FLOP is around 1e11 OP or FLOP.

When inference compute is higher than these levels, inference compute is 'above compute optimality.' For example, if an AI developer increases the inference compute from 1e11 OP or FLOP to 1e14 OP or FLOP, there is an increase of 3 orders of magnitude ('OOM') above the compute-optimal inference compute.[90]

---

[84] *See* Lennart Heim et al., *Governing Through the Cloud: The Intermediary Role of Compute Providers in AI Regulation*, ARXIV (Mar. 2024), https://doi.org/10.48550/arXiv.2403.08501, at 23.

[85] Ning Chen et al., *Towards Integrated Fine-tuning and Inference when Generative AI meets Edge Intelligence*, ARXIV (Jan. 5, 2024), https://doi.org/10.48550/arXiv.2401.02668, at 3 ("Terminals input unlabeled data (e.g., prompt instruction and images, etc.) into the pre-trained and fine-tuned GAI model, and the GAI model outputs the content following the user intention, such as reply text and drawn images, so as to meet users' service requirements.").

[86] The 'per request' specification is useful to distinguish the compute used with *each* inference request from users from the cumulative compute usage for inference, i.e., the inference compute necessary to serve *all* users. *See* Pistillo et al., supra note 1, at 8-9 and 21.

[87] *See* Hoffmann et al., *supra* note 47; Nikhil Sardana et al., *Beyond Chinchilla-Optimal: Accounting for Inference in Language Model Scaling Laws*, ARXIV (last updated July 18, 2024), https://doi.org/10.48550/arXiv.2401.00448.

[88] Compute optimality is a somewhat artificial optimality criterion and most models actually deviate from it for engineering or economic reasons. Nevertheless, it serves as a useful baseline to make comparisons.

[89] *See* Hoffmann et al., *supra* note 47; Nikhil Sardana et al., *Beyond Chinchilla-Optimal: Accounting for Inference in Language Model Scaling Laws*, ARXIV (last updated July 18, 2024), https://doi.org/10.48550/arXiv.2401.00448.

[90] An OOM is a factor of 10. Therefore, one OOM decreases the floating number by one unit, e.g., from 10^26 to 10^25. 2 OOM equals a factor of 100.

*(ii)* *Potential Loophole*

There is a proven relationship between above compute-optimal inference compute and capabilities, as shown in Figure 7 below.[91] Specifically, an increase of 2-3 OOM above compute-optimal inference compute is estimated to allow developers to save ~2 OOM in training compute, while maintaining performance.[92] Capabilities usually plateau after that, with the exception of math and coding. In the latter case, an increase of 5-6 OOM above compute-optimal inference compute corresponds to a saving of 3-4 OOM in training compute.[93]

Figure 7 – Relationship Between Inference and Capabilities (General, and Math and Coding)

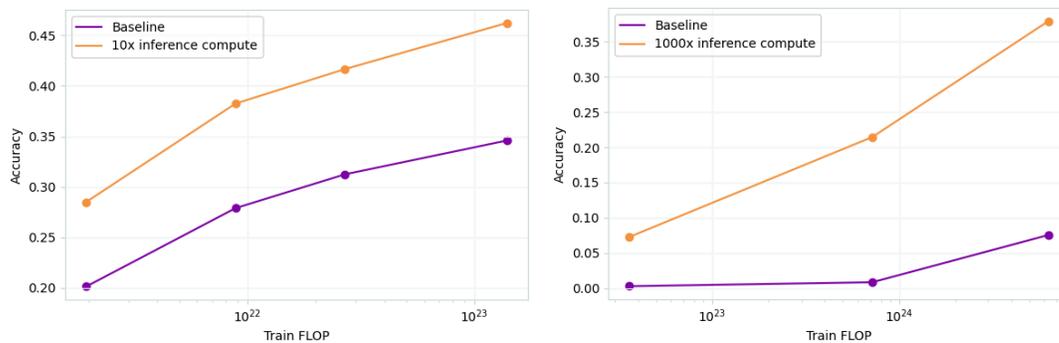

Performance of a code model with best-of-n sampling. Left: in a math benchmark, increasing the number of samples by a factor of 10 produces improvements as significant as increasing training compute by a factor of 5.[94] Right: in a coding benchmark, an increase of 1000x in inference compute produces savings comparable to more than two orders of magnitude of training compute scaling.[95]

Within these limits, it is possible to increase a model's capabilities at the cost of additional inference compute per request.[96] Hence, AI developers could shift compute from training to inference,[97] thus obtaining equally-capable systems while utilizing less training compute.[98] This would allow developers to stay within established training compute thresholds while obtaining more capable AI models, as shown in Figure 8 below.

---

[91] *See* Pablo Villalobos & David Atkinson, *Trading Off Compute in Training and Inference,* EPOCH AI (July 28, 2023), https://epochai.org/blog/trading-off-compute-in-training-and-inference.
[92] *Id.*
[93] *Id.*
[94] Data from Yujia Li et al., *Competition-Level Code Generation with AlphaCode*, ARXIV (Feb. 8, 2022), https://doi.org/10.1126/science.abq1158.
[95] Data from Bradley Brown et al., *Large Language Monkeys: Scaling Inference Compute with Repeated Sampling*, ARXIV (last updated Sep. 16, 2024), https://doi.org/10.48550/arXiv.2407.21787.
[96] *Id. See also* Pistillo et al., supra note 1, at 22.
[97] Recent developments in AI—such as the release of OpenAI o1, a model that specifically relies on this technique—suggest that this avenue of improvement might become more relevant in the future.
[98] *See* Hooker, *supra* note 2, at 13 (compute "can be grouped under 'inference-time compute' and can result in large performance gains that dramatically increase the risk profile of a model.").

Figure 8 – Inference as a Potential Loophole

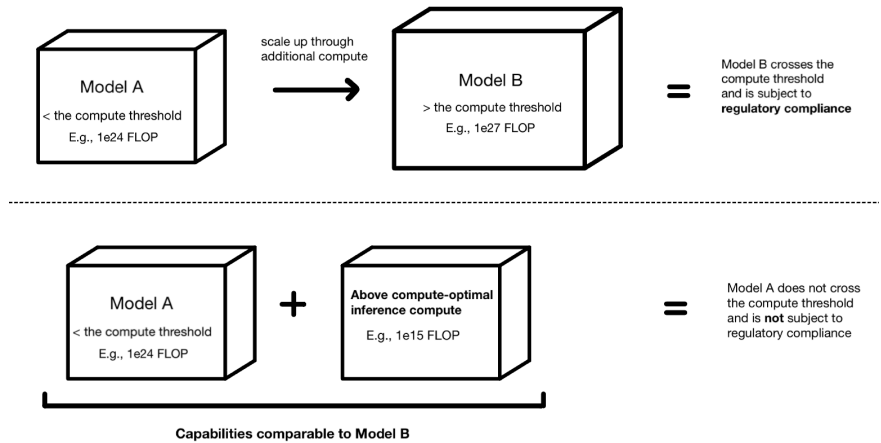

*(iii)* *Legal Frameworks*

**(a)** *Executive Order 14110*
Inference compute is not considered by Executive Order 14110. Therefore, it remains unclear if and how inference compute will be considered for the purposes of identifying covered models, and AI developers could exploit this gap to game training compute thresholds (*see* Figure 8 above).

**(b)** *AI Act*
The AI Act also does not appear to take into account inference compute in the cumulative training compute threshold, opening the door to the potential loophole described above (*see* Figure 8 above).

**(c)** *California Senate Bill 1047*
Similarly, inference compute was not considered in the vetoed California Senate Bill 1047 (*see* Figure 8 above).

*(iv)* *Policy Recommendation*

As AI models continue to scale, we recommend that **above-optimal inference compute should be considered when calculating training compute thresholds**. In particular, **training compute thresholds should be deemed as crossed if the above the compute-optimal inference** **compute is equal to an increase in training compute that would have caused the AI model to cross the training compute threshold**.

Consider for example[99] a 1e24 OP or FLOP model, whose compute-optimal inference compute sits around 1e11 OP or FLOP. If an AI developer uses 1e15 OP or FLOP instead of 1e11 OP or FLOP, the amount of inference compute 'in excess' (+3 OOM, equivalent to ~2 OOM in training compute) should be added to the training compute used to develop the model. In this example, the total would be 1e26 FLOP or OP (1e24 FLOP or OP + 2 OOM), and the AI developer would have crossed the compute threshold in the European Union.

This recommendation encounters two main uncertainties. First, trying to bypass compute thresholds using above-optimal inference compute could be economically inconvenient for AI developers. Increasing inference compute harms the scalability of AI systems, as each interaction with the system becomes more computationally expensive. Hence, it is possible that this technique will pose less concerns than expected. Second, enforcing this type of policy approach could be challenging for math and coding, where an increase of 5-6 OOM above compute-optimal inference compute corresponds to a saving of 3-4 OOM in training compute. There are hundreds of models above 1e23 OP or FLOP, and even more above 1e22 OP or FLOP,[100] making it difficult for regulators to monitor above-optimal inference compute use.

### III.  Conclusion

Despite the criticisms reasonably leveled at training compute thresholds, this form of regulatory threshold is relied on in existing legal frameworks, such as Executive Order 14110 and the AI Act and, and might appear in upcoming frameworks as well, such as the UK frontier AI Bill. The main reason for its use in AI legal frameworks is the ease with which it can, albeit imperfectly, pre-identify potentially dangerous systems of concern and trigger additional regulatory attention.

Considering the importance of training compute thresholds for AI governance at the time of writing, in this paper we examined some potential legal loopholes to compute thresholds, assessed how they interact with existing legal frameworks, and recommended policy solutions to increase the robustness of this regulatory tool.

---

[99] This example is purely illustrative. The equivalence between training compute and inference compute could change in different tasks and applications.

[100] *See Large-Scale AI Models*, *supra* note 12 (last consulted Dec. 30, 2024).

In summary:

I. On **fine-tuning**:
   - We observed that:
     - With 15% of fine-tuning compute, AI developers could generate new general capabilities.
     - Fine-tuning is not taken into account by the compute threshold set forth under Executive Order 14110. This could constitute a legal loophole as developers could increase model capabilities through intensive fine-tuning without incurring regulatory attention.
   - We recommended that, when the fine-tuning compute exceeds 15% of the original training compute (in one instance or the aggregate), the United States calculate fine-tuning compute as counting towards meeting the 1e26 OP or FLOP compute threshold established under Executive Order 14110.

II. On **model reuse**:
   - We observed that:
     - Through model reuse, AI developers can use up to ~10x less compute to develop an AI model. This constitutes a loophole because the resulting model would not fall within the compute threshold while having capabilities comparable to the starting model.
     - Model reuse is not addressed by Executive Order 14110 or the AI Act.
   - Hence, we recommended that the United States and the European Union take into account the up to ~10x compute savings associated with model reuse when calculating the compute threshold for reused models. Countries could also subject the starting model to compute thresholds if such models independently cross established compute thresholds, whether incognito or not.

III. On **model expansion**:
   - We observed that:
     - This technique allows compute savings in the range of 20-76%.
     - Executive Order 14110 and the AI Act do not address model expansion.
   - Hence, we recommended that the United States and the European Union take into account the 20-76% compute savings associated with model expansion when calculating the compute threshold, and subject an AI model to regulatory compliance when it results from model expansion and crosses 5e25 OP or FLOP in the United States, or 5e24 FLOP in the European Union.

IV. On **above compute-optimal inference compute**:
- We observed that:
    - An increase of 2-3 OOM above compute-optimal inference compute usually enables a saving of ~2 OOM in training compute (5-6 OOM for a saving of 3-4 OOM, for math and coding capabilities).
    - Executive Order 14110 or the AI Act do not consider inference compute.
- Hence, we recommended that the United States and the European Union consider training compute thresholds as crossed if the above the compute-optimal inference compute is equal to an increase in training compute that would have caused the AI model to cross the training compute threshold.

# Acknowledgments

We thank Charlotte Stix, Jaime Sevilla, Alexander Erben, Cullen O'Keefe, and Matthijs Maas for thoughts and comments to this paper. Mistakes are our own.